\documentclass[fleqn,twoside]{article}
\usepackage{espcrc2,epsfig}

\def\be{\begin{equation}}
\def\ee{\end{equation}}
\def\bea{\begin{eqnarray}}
\def\eea{\end{eqnarray}}
\def\nn{\nonumber}

\newcommand{\rhobar}{\bar {\rho}}
\newcommand{\etabar}{\bar{\eta}}
\newcommand{\epsilonk}{\left|\varepsilon_K \right|}
\newcommand{\vubovcb}{\left | \frac{V_{ub}}{V_{cb}} \right |}
\newcommand{\vubsvcb}{\left | V_{ub}/V_{cb}  \right |}

\newcommand{\dmd}{\Delta m_d}
\newcommand{\dms}{\Delta m_s}
\newcommand{\GeV}{\rm{GeV}}
\newcommand{\MeV}{\rm MeV}
\newcommand{\BK}{B_K}
\newcommand{\Bd}{ {B}^{0}_{d} }
\newcommand{\Bdb}{ \bar {B}^{0}_{d}  }
\newcommand{\Bs}{ B^0_s }
\newcommand{\Bsb}{ \bar {B}^0_s }
\newcommand{\fbdsqbd}{f_{B_d} \sqrt{\hat B_{B_d}}}

\newcommand{\Vcb}{\left | { V}_{cb} \right |}
\newcommand{\Vub}{\left | { V}_{ub} \right |}

\title{Status of the CKM matrix}

\author{M. Ciuchini\address{Dip. di Fisica, Univ. di Roma Tre and INFN Sez. di Roma III,\\
Via della Vasca Navale 84, I-00146 Roma (Italy)}}

\begin{document}

\begin{abstract}
An updated determination of the parameters of the Cabibbo-Kobayashi-Maskawa matrix is presented.
\vspace{1pc}
\end{abstract}

\maketitle
\section{Introduction}
In the Standard Model,  weak interactions of quarks are governed
by the four parameters of the CKM matrix~\cite{ref:ckm} which,
in the Wolfenstein-Buras  parametrisation~\cite{ref:wolf,ref:blo},
are labelled as $\lambda,~A,~\rhobar~{\rm and}~\etabar$.
Measurements of semileptonic decays of strange and beauty particles
are the main sources of information on $\lambda$ and $A$, respectively.
The values of $\epsilonk$, $\vubsvcb$, $\dmd$ and ${\dms}$
provide a set of four constraints for $\rhobar$ and $\etabar$.
These constraints depend, in addition, on other quantities obtained
from measurements and/or theoretical calculations.
The regions of  $\rhobar$ and $\etabar$ preferred by the four constraints
are expected to overlap, as long as the Standard Model
gives an overall description of the various experimental observations.
In this paper, we summarize the results of our analysis of
the
CKM matrix. Further details on various aspects of this analysis can be found
in ref.~\cite{ref:ckmnoi}.

\section{Basic Formulae}
Four measurements restrict, at present, the possible range of variations of
the $\rhobar$
and $\etabar$ parameters:
\begin{itemize}
\item The relative rate of charmed and charmless $b$-hadron semileptonic
decays
which allows to measure the ratio
\begin{equation}
\vubovcb = \frac{\lambda}{1~-~\frac{\lambda^2}{2}}\sqrt{\rhobar^2+\etabar^2}\, .
\label{eq:C_vubovcb} \end{equation}.
\item The ${B}^0_d-\bar{{B}}^0_d$ time oscillation period
which can be related to the mass difference between the light and heavy
mass eigenstates
of the ${B}^0_d-\bar{{B}}^0_d$ system
\bea
\dmd\ &=&
\ {G_F^2\over 6 \pi^2} m_W^2 \ \eta_c S(x_t) \ A^2 \lambda^6 \ [(1-\nn\\
&&\rhobar)^2+\etabar^2] \ m_{B_d} \ f_{B_d}^2 \hat B_{B_d} \ ,
\label{eq:deltam}
\eea
where $S(x_t)$ is the Inami-Lim function~\cite{inami} and  $x_t=m_t^2/M^2_W$.
$m_t$ is the $\overline{MS}$ top mass,
$m_t^{\overline{MS}}(m_t^{\overline{MS}})$,  and
$\eta_c$ is the perturbative QCD short-distance NLO correction.
The remaining factor, $f_{B_d}^2 \hat B_{B_d}$, encodes the information
of non-perturbative QCD.
Apart for $\rhobar$ and $\etabar$, the most uncertain parameter in this
expression is $f_{B_d} \sqrt{\hat B_{B_d}}$. The value of $\eta_c=0.55 \pm 0.01$
has been obtained in~\cite{ref:bur1} and
we used   $m_t=(167 \pm 5)\,  \GeV$, as
deduced from measurements of the  mass by CDF and D0 Collaborations
\cite{ref:top}.
\item
The limit on the lower value for the time oscillation period
of the ${B}^0_s-\bar{{B}}^0_s$ system is transformed
into a limit on $\dms$ and compared with $\dmd$
\bea
\frac{\dmd}{\dms}&=&\frac{m_{B_d}f_{B_d}^2 \hat B_{B_d}}
{m_{B_s}f_{B_s}^2 \hat B_{B_s} }\
\left( \frac{\lambda}{1-\frac{\lambda^2}{2}} \right )^2 \times\nn\\
&& [(1-\rhobar)^2+\etabar^2]\, .
\label{eq:dms}
\eea
The ratio
$\xi~=~f_{B_s}\sqrt{\hat B_{B_s}}/f_{B_d}\sqrt{ \hat B_{B_d}}$
is expected to be better determined from theory than the individual
quantities entering into its expression. In  our analysis, we accounted for the
correlation due to the appearance of $\dmd$ in both Equations~(\ref{eq:deltam})
and (\ref{eq:dms}).
\item CP violation in the kaon system which is expressed by $\epsilonk$
\bea
\epsilonk\ &=& C_\varepsilon \ A^2 \lambda^6 \ \etabar
\Big[ -\eta_1 S(x_c)+\nn\\
&& \eta_2 S(x_t) \left( A^2 \lambda ^4 \left(1-\rhobar\right)\right)+\nn\\
&& \eta_3 S(x_c,x_t) \Big] \ \hat B_K\ ,
\label{eq:epskdef}
\eea
where
\begin{equation} C_\varepsilon = \frac{G_F^2 f_K^2 m_K m_W^2}
{6 \sqrt{2}
\pi^2 \Delta m_K} \ . \end{equation}
$S(x_i)$ and $S(x_i,x_j)$ are the appropriate
Inami-Lim functions~\cite{inami} of $x_q=m_q^2/m_W^2$, including the
next-to-leading order QCD corrections~\cite{ref:bur1,ref:basics}.
The most uncertain parameter is $\hat \BK$.
\end{itemize}
Constraints are obtained by comparing present measurements with
theoretical expectations using the expressions given above
and taking into account the different sources of uncertainties. In addition
to $\rhobar$ and $\etabar$, these expressions depend on other quantities
which have been listed in Table \ref{tab:1}.
Additional measurements or theoretical
determinations have been used to provide information on the values of
 these parameters.

\begin{table*}[htb]
\caption{Values of the quantities entering into the expressions
of $\epsilonk$, $\vubsvcb$, $\dmd$ and $\dms$.
In the third and fourth columns the Gaussian and the flat part of the
uncertainty are given, respectively.}
\label{tab:1}
\begin{center}
\begin{tabular}{|c|c|c|c|r|}
\hline
                       Parameter                 &  Value &
 Gaussian &  Uniform & Ref. \\
 & &   $\sigma$ & half-width &  \\
\hline
$\lambda$ & $0.2237$ & \multicolumn{2}{|c|}{  $0.0033$} & \cite{ref:ckmnoi} \\
$\left | V_{cb} \right |$ & $40.7 \times 10^{-3}$ &
\multicolumn{2}{|c|}{$1.9\times 10^{-3}$} & \cite{ref:ckmnoi}\\
$\left | V_{ub} \right |$  & $ 36.1  \times 10^{-4}$ & $ 4.6 \times 10^{-4}$ & --
&\cite{ref:ckmnoi}\\
$\epsilonk$   &
$2.271 \times 10^{-3}$&
$0.017
 \times 10^{-3}$  & --
     &\cite{ref:pdg00} \\
     $\Delta m_d$  & $0.489~\mbox{ps}^{-1}$ & $0.008~\mbox{ps}^{-1}$ & --
     &\cite{ref:osciw}  \\
$\Delta m_s$  & $>$ 14.6 ps$^{-1}$ at 95\% C.L.
     & \multicolumn{2}{|c|}{see text} &~\cite{ref:osciw}  \\
$m_t$ & $167~\GeV$ & $ 5~\GeV$ & --
     &\cite{ref:top} \\
$m_b$ & $4.23~\GeV$ & $ 0.07~\GeV$ & --
     &\cite{ref:hispanicus} \\
$m_c$  & $1.3~\GeV $ & $0.1~\GeV$
     & --  &\cite{ref:pdg00} \\
$\hat \BK$    & $0.87$ & $0.06$ &  $0.13$
     &   \cite{ref:ckmnoi}   \\
$f_{B_d} \sqrt{\hat B_{B_d}}$ & $230~\MeV$  & $25~\MeV$
     &  $20~\MeV$  & \cite{ref:ckmnoi} \\
$\xi=\frac{ f_{B_s}\sqrt{\hat B_{B_s}}}{ f_{B_d}\sqrt{\hat B_{B_d}}}$
     & $1.14$ & 0.04 & $ 0.05 $& \cite{ref:ckmnoi} \\
$\alpha_s$  &  $0.119$ & $0.003$ & -- &\cite{ref:basics} \\
$\eta_1$  &  $1.38$ & $0.53$ & -- &\cite{ref:basics} \\
$\eta_2$  &  $0.574$ & $ 0.004$ & --
     &\cite{ref:bur1} \\
$\eta_3$  &  $0.47$ & $ 0.04$ & -- &\cite{ref:basics} \\
$\eta_b$  &  $0.55$ & $ 0.01$ & --
     &\cite{ref:bur1} \\
$f_K$     & $0.159~\GeV$ & \multicolumn{2}{|c|}{fixed}
     &\cite{ref:pdg00} \\
$\Delta m_K$  & $
0.5301
\times 10^{-2} ~\mbox{ps}^{-1}$
     & \multicolumn{2}{|c|}{fixed} &~\cite{ref:pdg00} \\
 $G_F $   & $
 1.16639
\times 10^{-5}~\GeV^{-2}$
     & \multicolumn{2}{|c|}{fixed}&~\cite{ref:pdg00} \\
$ m_{W}$  & $80.42
 ~\GeV $
     & \multicolumn{2}{|c|}{fixed} & ~\cite{ref:pdg00} \\
$ m_{B^0_d}$ & $5.2792
~\GeV $
     & \multicolumn{2}{|c|}{fixed} &~\cite{ref:pdg00} \\
$ m_{B^0_s}$ & $5.3693
~\GeV $
     & \multicolumn{2}{|c|}{fixed} &~\cite{ref:pdg00} \\
$ m_K$ & $0.493677
 ~\GeV$
     & \multicolumn{2}{|c|}{fixed} &~\cite{ref:pdg00} \\
\hline
\end{tabular}
\end{center}
\end{table*}

\section{Inferential framework}
The phenomenological analysis is performed using the Bayesian inference.
In this framework, every parameter entering the constraints, regardless
their theoretical or experimental origin, is characterized by a
probability density function (p.d.f.). We assign the p.d.f.s of the different
parameters as shown in Table \ref{tab:1}. These distributions are taken as Gaussian
or flat or a convolution of the two, according to the origin
of the uncertainty being purely statistical or coming from influence quantities
(such as theoretical parameters or systematic errors in experiments). The
parameters $\rhobar$ and $\etabar$ have also an a-priori p.d.f. which is assumed
to be flat. Using these distributions and the experimental constraints discussed
in the previous section, we can build an overall likelihood and obtain
a-posteriori p.d.f.s for $\rhobar$ and $\etabar$ or any other quantities of
interest.

This method provides a theoretically-sound approach that allows a consistent treatment of
the systematic and theoretical uncertainties and makes it possible to define regions
where the values of $\rhobar$ and $\etabar$ (as well as of other quantities) are contained
with any given level of probability.

The Bayesian method applied to the CKM-matrix analysis is discussed at length in
ref.~\cite{ref:ckmnoi}.

A different analysis based on frequentistic techniques
can be found in ref.~\cite{Hocker:2001xe}. This paper also puts forth a rather academic
argument that, in the mind of its authors, should demonstrate that the Bayesian
method unfairly narrows the region of predicted results for quantities depending
on more than one theoretical parameter. The argument is the following:
consider a quantity $T_P=x_1 x_2 x_3 \dots x_N$,
defined as the product of $N$ ``theoretical'' parameters $x_i$, and assume to know
that these parameters lie in the range $[-1, 1]$. Following the Bayesian method
one easily finds that the resulting p.d.f. of $T_P$ peaks
at zero and becomes more and more peaked as the number $N$ of parameters
increases. This leads the authors of refs.~\cite{Hocker:2001xe,Stone:2001jh} to conclude
that the Bayesian method is ``dangerous'', since a safe method should
have predicted $T_P$ in the range $[-1, 1]$. However, our conclusion is
quite the opposite. It is perfectly natural that the p.d.f. of $T_P$ peaks as $N$ increases,
being simply the effect of combinatorics (unless, of course, there are reasons to believe
that the different determinations of the $x_i$ are correlated).
The singularity of the p.d.f. as $N$ goes to infinity, which has been pointed out as
a pathology of the Bayesian method, has also a simple explanation: the
knowledge of an infinite number of p.d.f. for the $x_i$ corresponds to determining $T_P$
with infinite precision. Therefore we believe that the argument presented in ref.~\cite{Hocker:2001xe}
does not affect the validity of the Bayesian method in any way. We rather find that methods
which predict $T_P$ in the range $[-1, 1]$ are deliberately throwing away
information. This may be justified and reasonable in specific and well-motivated cases, but as
a general rule, we find it quite contrary to the spirit of this kind of analyses aiming to use present
theoretical and experimental information to extract the best determination of the unitarity triangle
in the Standard Model.

\begin{figure}[t]
\epsfig{figure=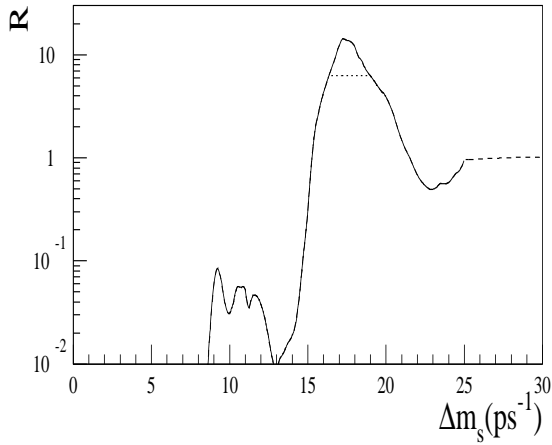,width=0.45 \textwidth,height=0.3 \textheight}
\caption{The likelihood ratio ${\rm R}(\dms)$ used in the analysis.}
\label{fig:pdf}
\end{figure}

\section{Input Parameters}
\label{inputs}
The values of the input parameters used in the anaysis are collected in Table
\ref{tab:1}. For all the theoretical parameters, we have used results
taken from lattice QCD. There are several reasons for this choice, which has
been adopted also in previous studies of the unitarity
triangle~\cite{ref:pion1,ref:pion4,ref:pprs,ref:parodietal,ref:roma}.
Lattice QCD is not a model,
as the  quark model for example, and therefore physical quantities can be
computed from first principles without arbitrary assumptions.
It provides a method for predicting all physical quantities
(decay constants, weak amplitudes, form factors) within  a unique, coherent
theoretical framework. For many quantities the statistical
errors have been reduced to the percent level.
Although most of the results are affected by systematic effects,
the latter can be ``systematically'' studied and eventually corrected.
All the recent literature on lattice calculations is indeed focused on
discussions of the systematic errors and studies intended to reduce these
sources of uncertainty. Finally, in cases where predictions from lattice QCD
have been compared with experiments, for example
$f_{D_s}$, the agreement has been found  very good.
Obviously, for some quantities the uncertainty from lattice simulations
is far from being satisfactory and  further effort is needed
to improve the situation. Nevertheless, for the
reasons  mentioned before, we think  that lattice results and uncertainties
are the most reliable ones and  we have used them in our study.

Results from the LEP working groups have been used for $\dmd$ and $\dms$.
LEP and CLEO measurements of $\Vcb$ and $\Vub$ have been combined to
obtain the values in Table \ref{tab:1}. Details on the methods used for
combining the various measurements can be found in \cite{ref:ckmnoi}.
In order to include the information from the lower limit on $\dms$ in the
analysis, we have found it better to use the likelihood ratio R defined as
\be
{\rm R}(\dms) = e^{\textstyle {-\Delta \log {\cal L}^{\infty}(\dms)}}
= \frac{{\cal L}(\dms)}{{\cal L}(\infty)} \ ,
\label{R_eq}
\ee
with
\begin{eqnarray}
\Delta \log{\cal L}^{\infty}(\dms)
&=& \frac{\textstyle 1}{\textstyle 2}\,\left[ \left(\frac{\textstyle {\cal A}-
1 }{\textstyle \sigma_{\cal A} }\right)^2-
 \left(\frac{\textstyle {\cal A}}{\textstyle\sigma_{\cal A}}\right)^2 \right] \nn\\
&=& \left (\frac{\textstyle 1}{\textstyle 2}
-{\cal A}\right){\textstyle 1\over
   \textstyle \sigma_{\cal A}^2}\ ,
\label{dms_right3}
\end{eqnarray}
where ${\cal A}$ is the measured oscillation amplitude at a given value of
$\dms$, expected to be equal to one at the physical $\dms$ and to
vanish elsewhere. The likelihood ratio is more effective than the usual
likelihood as it exploits the fact that the oscillation is signaled by the
amplitude being {\em both} compatible with one {\em and} incompatible with zero.
A thorough discussion of this method is presented in \cite{ref:ckmnoi}.
The likelihood ratio R used in the analysis is shown in Figure \ref{fig:pdf}.

\begin{figure*}[htb]
\begin{tabular}{cc}
\vspace{-40pt} &\\
\multicolumn{2}{c}{
\epsfig{figure=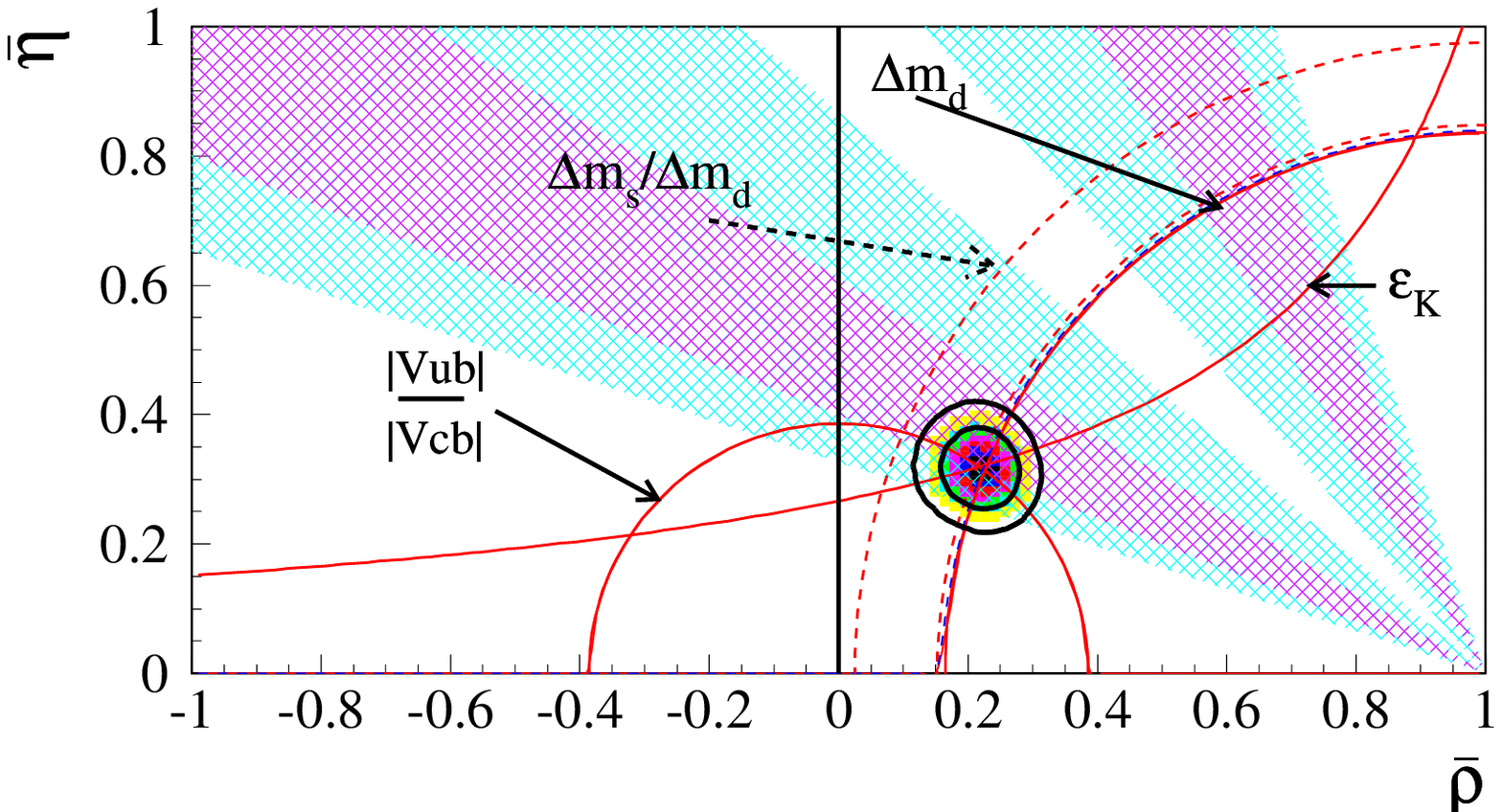,width=0.9\textwidth,
}} \\
\epsfig{figure=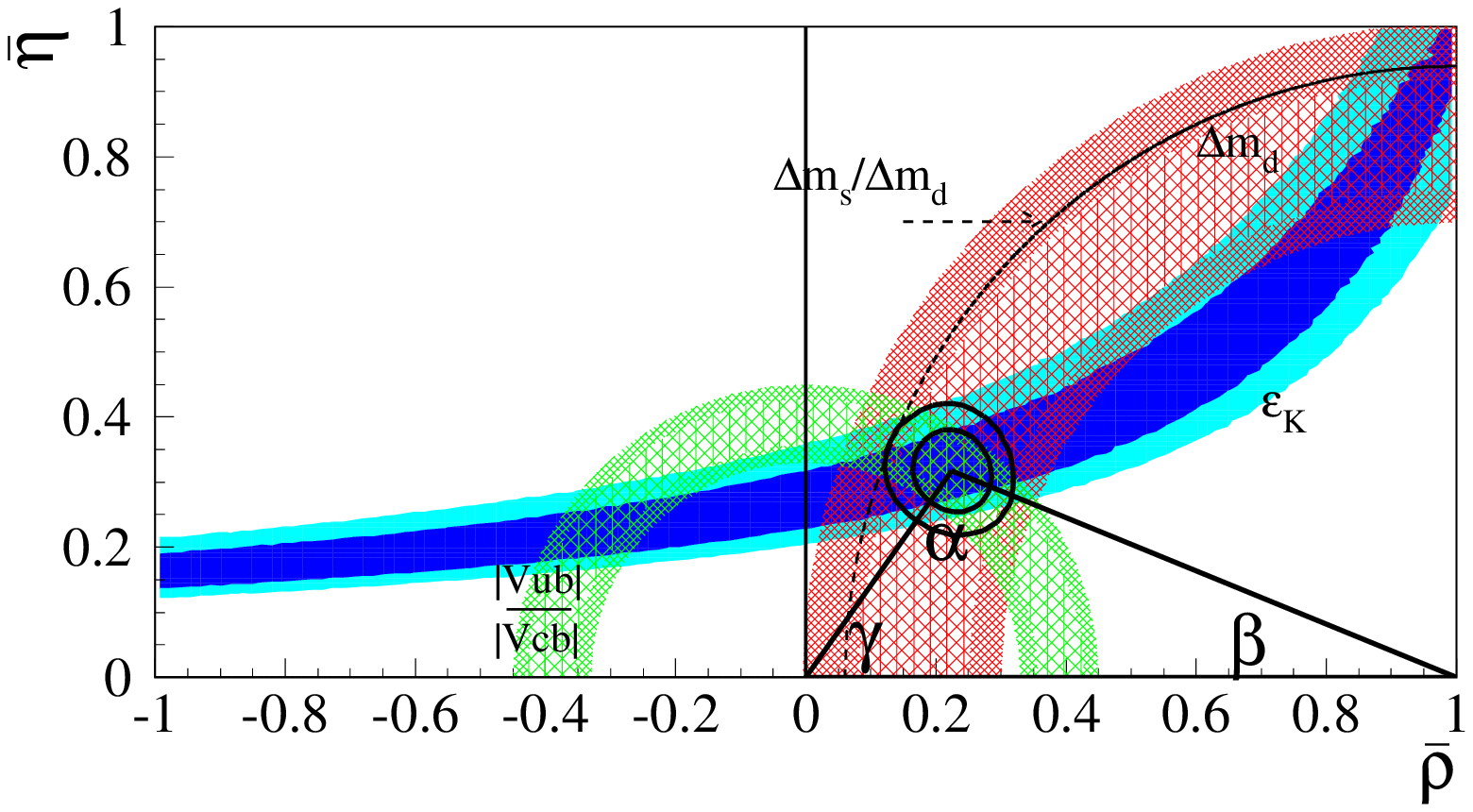,width=0.47\textwidth,
height=0.25\textheight} &
\epsfig{figure=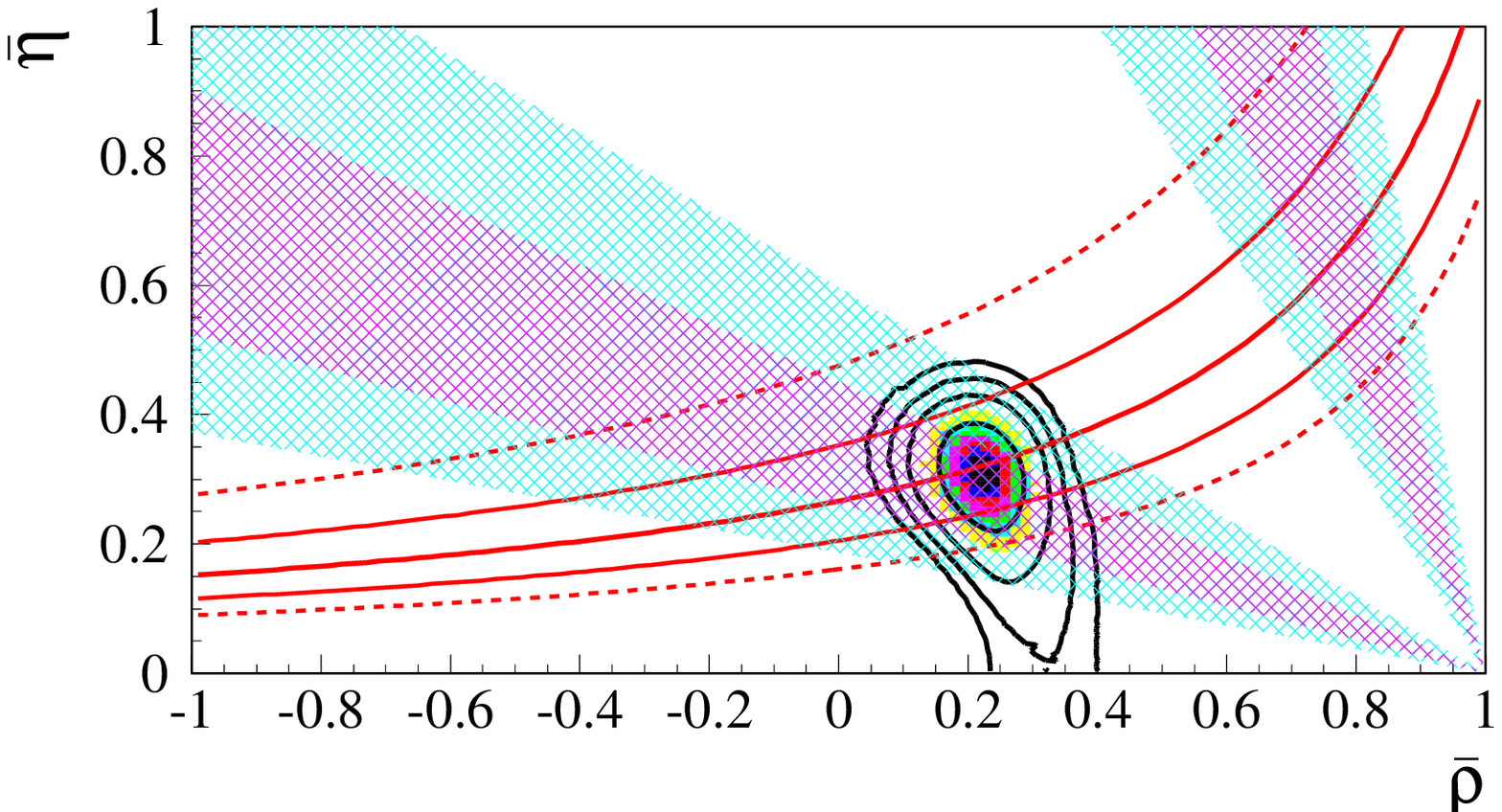,width=0.47\textwidth,
height=0.25\textheight}
\end{tabular}
\caption{Contour plots in the ($\rhobar$,$\etabar$) plane.}
\label{fig:contours}
\end{figure*}

\section{Results}
The region in the ($\rhobar,~\etabar)$ plane selected by the measurements
of $\epsilonk$, $\vubsvcb$, $\dmd$ and from the information
on $\dms$ (using the R function in Figure \ref{fig:pdf})
is given in the upper part of Figure~\ref{fig:contours}. On the lower-left
part, the uncertainty bands for the quantities, obtained using
Equations~(\ref{eq:C_vubovcb})--(\ref{eq:epskdef}), are presented. Each band,
corresponding to only  one of  the constraints,  contains 68\% and 95\%  of the
events obtained by varying  the input parameters. This comparison illustrates
the consistency of the different constraints provided by the Standard Model.
The measured values of $\rhobar$ and $\etabar$ are
\begin{equation}
\rhobar=0.218 \pm 0.038  ,~\etabar=0.316 \pm 0.040
\label{eq:eta1}
\end{equation}
The two quantities are practically uncorrelated (correlation coefficient of -5\%),
as it can be seen from the contour plots in the ($\rhobar,~\etabar)$ plane.
Fitted values for the angles of the unitarity triangle have been also obtained
\bea
&&\!\!\!\!\!\sin(2\,\beta)= 0.696 \pm 0.068\, ,~
\gamma=(55.5 \pm 6.2)^{\circ}\, , \nn\\
&&\!\!\!\!\!\sin(2\,\alpha)= -0.42 \pm  0.24\, .
\label{eq:eta2}
\eea

\begin{figure*}[t]
\begin{tabular}{cc}
\epsfig{figure=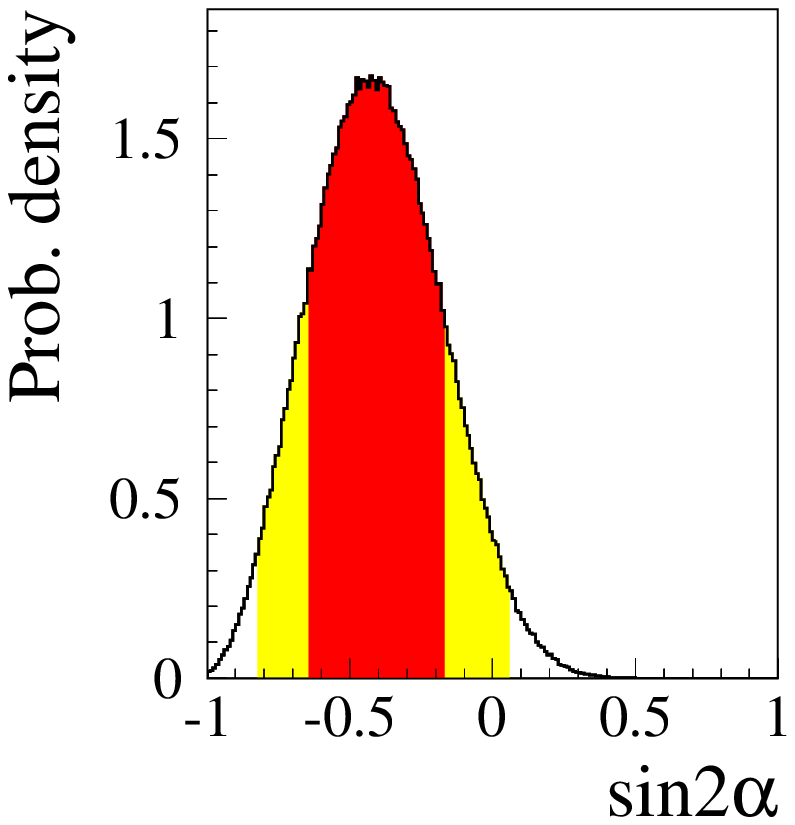,width=0.47 \textwidth,height=0.25
\textheight} &
\epsfig{figure=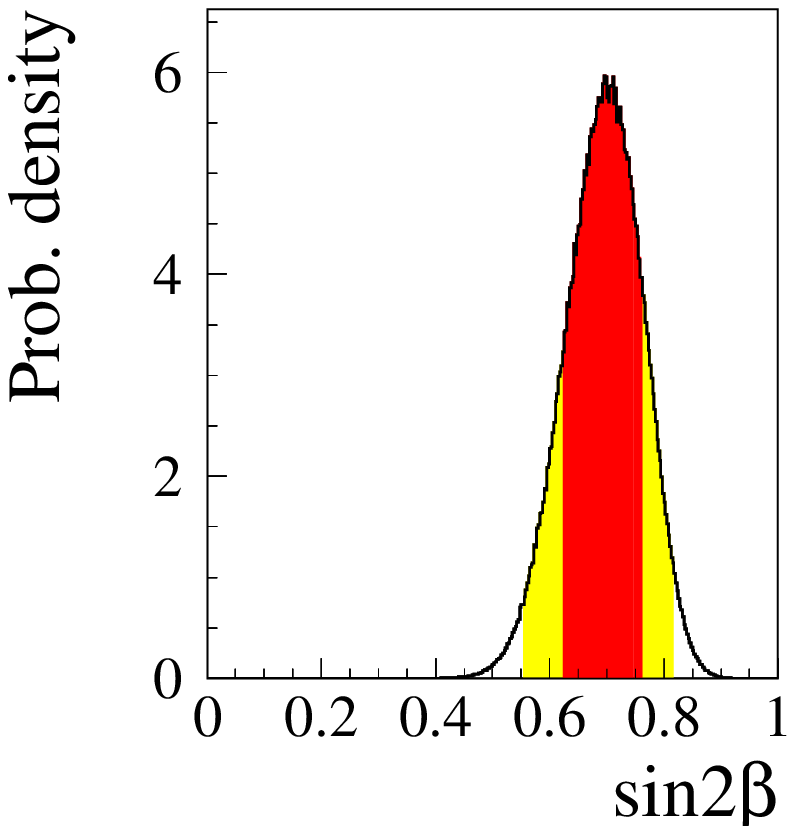,width=0.47 \textwidth,height=0.25
\textheight}\\
\epsfig{figure=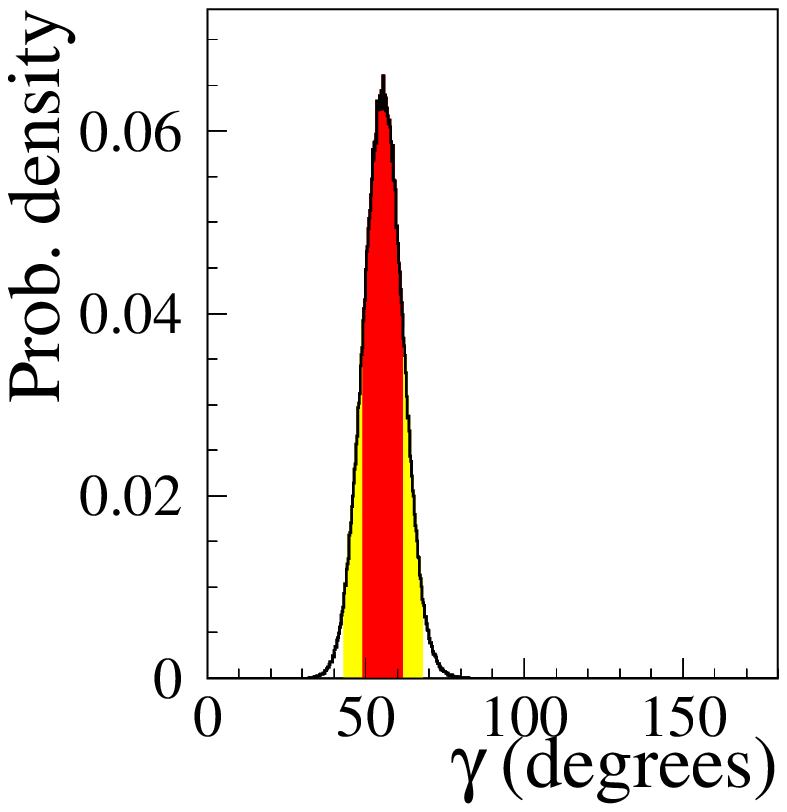,width=0.47
\textwidth,height=0.25 \textheight} &
\epsfig{figure=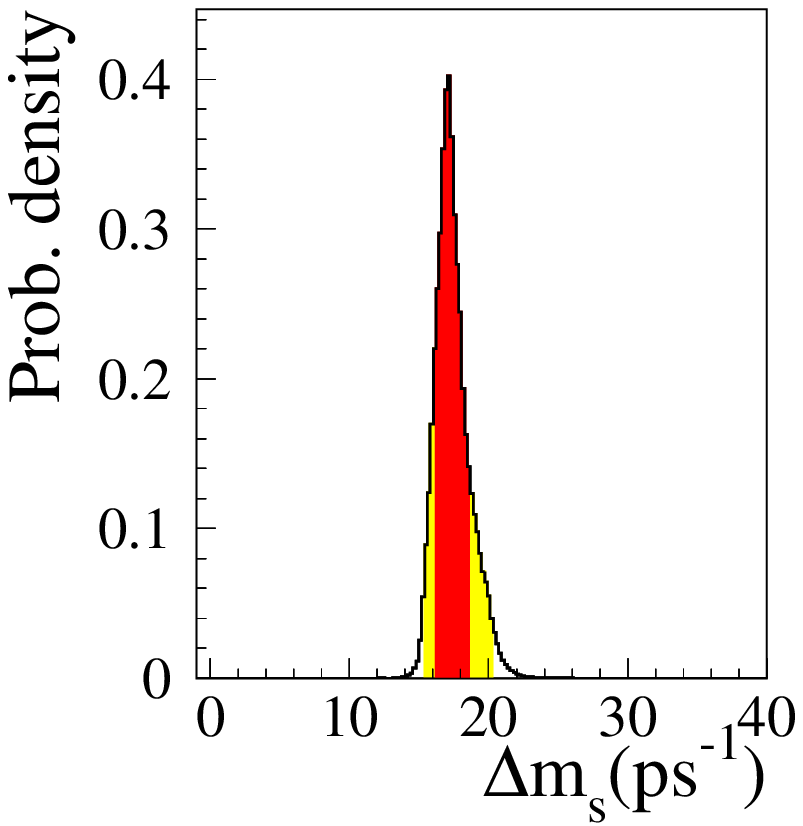,width=0.47
\textwidth,height=0.25 \textheight}
\end{tabular}
\caption{Some a-posteriori p.d.f.}
\label{fig:results}
 \end{figure*}

In Figure~\ref{fig:results}, the p.d.f. for the angles of
the unitarity triangle are given. Few comments can be made:
\begin{itemize}
\item
$\sin(2\beta)$ is determined quite accurately. This value has
to be compared with the world average $\sin(2\beta) = 0.79 \pm 0.10$~\cite{s2bwa};
\item
the angle $\gamma$ is known within an accuracy of about 10\%. It has to be stressed that,
with present measurements, the probability that
$\gamma$ is greater than 90$^{\circ}$ is only 0.03\%.
Without including the information from $\dms$, it is found that $\gamma$ has
4\% probability to be larger than 90$^{\circ}$.
\end{itemize}

As four constraints are used to determine the values of two parameters,
it is possible to relax, in turn,  one (or more)  of these constraints,
still obtaining significant confidence intervals.
An interesting exercise consists in  removing
the theoretical constraint for  $\hat B_K$ in  the measurement of $\epsilonk$
(\cite{ref:checchia}-\cite{ref:noepsk}).
The corresponding selected region in the ($\rhobar,~\etabar)$ plane is shown
in lower-right plot of Figure~\ref{fig:contours}, where the region selected by
the measurement of $\epsilonk$ alone is also drawn.
This comparison shows that the Standard Model picture of CP violation in the $K$ system and
of $B$ decays and oscillations are consistent.
In the same figure, we also compare the allowed regions in the ($\rhobar,~\etabar)$ plane
with those selected by the measurement of $\sin(2\,\beta)$ using $J/\psi K_S$ events.

Using constraints from $b$-physics alone the following results are obtained
\begin{equation}
\begin{array}{ll}
\etabar=0.304^{+0.050}_{-0.058}, &  \\
\qquad 0.167\le\etabar\le 0.400 & \rm{at}~95\%~{\rm prob.}\\
\sin(2\,\beta) = 0.676^{+0.078}_{-0.096}, &\\
\qquad 0.430\le\sin(2\,\beta)\le 0.820 & \rm{at}~95\%~{\rm prob.}  \\
\label{eq:eta32}
\end{array}
\end{equation}

Another way for illustrating the agreement between $K$ and $B$ measurements consists
in comparing the values of the $\hat \BK$ parameter obtained in lattice
QCD calculations with the value extracted  from
Equation~(\ref{eq:epskdef}), using the
values of $\rhobar$ and $\etabar$ selected by  $b$-physics alone
\begin{eqnarray}
&&\hat \BK^{\rm b-phys}=0.88^{+0.27}_{-0.13}  ,\nn\\
&&0.65 \le \hat \BK^{\rm b-phys} \le 1.65~{\rm at}~95\%~{\rm prob.}
\label{eq:bkb2}
\end{eqnarray}
Since $\hat \BK$ is not limited from above, for the present study, probabilities are
normalised assuming $\hat \BK<$5.

The importance of $\Bs$--$\Bsb$ mixing can be illustrated
from the p.d.f. of the angle $\gamma$ obtained with or without including the $\dms$
constraint (respectively light and dark lines in the lower-left plot of Figure
\ref{fig:results}). High values for $\gamma$ are excluded at high confidence
level by the experimental lower limit on $\dms$.

It is also possible to extract the probability distribution for $\dms$, from
which one obtains
\bea
&&\Delta m_s = (16.1 \pm 3.2) \ {\rm ps}^{-1}\,,\nn\\
&&9.4 \le  \Delta m_s   \le 23.0 ~{\rm ps}^{-1}~ {\rm at}~  95\% \ {\rm prob.}
\eea
If the information from the ${B}^0_s-\bar{{B}}^0_s$ analyses is included, results
become
\bea
&&\Delta m_s = (17.1^{+1.5}_{-0.9}) \ {\rm ps}^{-1}\, ,\nn\\
&&15.4 \le  \Delta m_s   \le 20.3 ~{\rm ps}^{-1}~{\rm at}~  95\% \
{\rm prob.}
\eea
These values are in agreement with the recent estimate of
$\Delta m_s = 15.8 (2.3) (3.3) \ ps^{-1}$, presented in~\cite{ape00}.

The value of $\fbdsqbd$ can be obtained by removing
the theoretical constraint coming from this parameter in $\Bd$--$\Bdb$ oscillations.
Using the two other theoretical inputs, $\hat \BK$
and $\xi$, $\fbdsqbd$ is measured with an accuracy
which is better than the current evaluation from lattice QCD, given
in Table \ref{tab:1}.  We obtain
\begin{equation}
\fbdsqbd = (228 \pm 12) \, \MeV \, .
\end{equation}
The present analysis shows that these results are in practice very weakly dependent
on the exact value taken for the uncertainty on $\fbdsqbd$.
An evaluation of this effect has been already presented in~\cite{ref:taipei}
where the flat part of the theoretical uncertainties on $\fbdsqbd$
was multiplied by two. Similar tests will be shown in Section \ref{sec:stability}.

It is even possible to remove the theoretical constraints on both
$f_{B_d}\sqrt{\hat B_{B_d}}$ and $\hat B_K$ and obtain the simultaneous lower bound
\begin{equation}
\hat B_K > 0.5\, ~~ {\rm and} ~~ f_{B_d}\sqrt{\hat B_{B_d}} >150\,\,\mbox{MeV}
\end{equation}
at $95\%$ probability.

\section{Stability of the results}
\label{sec:stability}
The sensitivity of present results on the assumed probability distributions attached
to the input parameters was studied.
The comparison of the results obtained by varying the size of the
theoretical  uncertainties
has been done to evaluate the sensitivity to these variations of uncertainties quoted
on fitted values. {\it This must not be taken as a proposal to inflate
the uncertainties obtained in the present analysis}.

In these tests, all values for uncertainties of theoretical origin have been, in turn,
multiplied by two. For the quantities $\Vub$ and $\Vcb$, new p.d.f. have been
determined, following the prescriptions mentioned in Section \ref{inputs}
and used in the analysis.
The main conclusion of this exercise is that, even in the case where all theoretical uncertainties are doubled,
the unitarity triangle parameters are determined with an uncertainty which increases only by about 1.5.

\begin{figure}[t]
\epsfig{figure=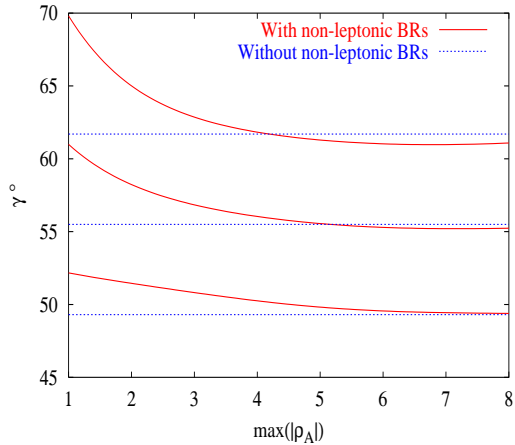,width=0.45 \textwidth,height=0.3 \textheight}
\caption{Dependence of $\gamma$ on the maximum value of the phenomenological
parameter $\vert\rho_A\vert$ appearing in the model used to compute $BR(B\to K\pi, \pi\pi)$. The bands
correspond to $1\sigma$ ranges.}
\label{fig:bbnsg}
\end{figure}

\section{CKM angle $\gamma$  from $B\to K\pi, \pi\pi$ decays}
It has been recently suggested that it is possible either to extract the CKM angle $\gamma$
using the measured $BR$s of $B\to K\pi, \pi\pi$ decays or, at least, use this experimental
information to improve the results of the unitarity triangle analysis~\cite{Beneke:2001ev}.
Although in principle this is certainly true, in practice it requires a very good control on the
hadronic uncertainties entering the theoretical predictions of these $BR$s. Contrary to recent
claims, however, the theoretical progresses in understanding in the infinite mass limit factorization of hadronic matrix
elements do not help in the specific case. In fact,  $B\to K\pi, \pi\pi$
channels get large, if not dominant, contributions from power-suppressed terms, for which a theory
has not been developed so far. For this reason, the approach of ref.~\cite{Beneke:2001ev}, where
power-suppressed contributions are factorized together with the leading terms, should be regarded as a phenomenological
model, and one with quite specific assumptions.  As a tool to extract fundamental parameters such as the
CKM angles, it has the same difficulties in assessing the theoretical uncertainties as any other model and therefore
should be used with caution.

To further illustrate this point, we have included in our analysis the constraint coming from $BR(B\to K\pi, \pi\pi)$
using the model of ref.~\cite{Beneke:2001ev} and considered the dependence of the predicted value
of $\gamma$ on the phenomenological parameter $\rho_A$.
This free parameter is introduced in the model to account to some extent for the infrared divergences
appearing in the perturbative calculation of power-suppressed corrections. According to ref.~\cite{Beneke:2001ev},
the allowed range of variation of $\rho_A$ is $\vert \rho_A\vert < 1$. We find instead that the data prefer
larger values $\vert\rho_A\vert\sim 4$--$6$.
Figure~\ref{fig:bbnsg} shows that the value of $\gamma$, which we obtain by including constraints from non-leptonic
$BR$s, deviates from the one of eq.~(\ref{eq:eta2}) only for value of max$(\vert\rho_A\vert)\sim 1$--$3$.
On the other hand, when
$\rho_A$ is allowed to get the values preferred by the data, no new informations on $\gamma$ are obtained
besides those coming from more reliable constraints.  In other words, it is just the restricted range for $\rho_A$ adopted in
ref. ~\cite{Beneke:2001ev} which originates the claimed ``improvement'' in the determination of the CKM parameters
due to non-leptonic $BR$s. To our knowledge, this choice has no compelling theoretical or
phenomenological basis.

We rather believe that, at present, our control of the hadronic uncertainties in the
$B\to K\pi, \pi\pi$ $BR$s is not developed enough to allow using these modes for costraining the CKM parameters.

\section*{Acknowledgements}
The results presented in this paper are based on work done in collaboration with
G. D'Agostini, E. Franco, V. Lubicz, G. Martinelli, F. Parodi, P. Roudeau, L. Silvestrini and
A. Stocchi.

\end{document}